\documentclass[12pt]{article}
\usepackage{amssymb,amsfonts}
\usepackage{graphics,psboxit,amsmath}
\usepackage{subfigure}
\usepackage{graphicx}
\usepackage{verbatim}
\usepackage[text={16.5cm,22.5cm},centering]{geometry}

\usepackage{color}
\usepackage{hyperref}
\hypersetup{colorlinks}

\definecolor{darkred}{rgb}{1,0,0}
\definecolor{darkgreen}{rgb}{0,0.5,0}
\definecolor{darkblue}{rgb}{0,0,1}
\definecolor{orange}{rgb}{1,0.5,0}
\definecolor{green}{rgb}{0,1,0}
\definecolor{purple}{rgb}{.5,0,1}

\hypersetup{ colorlinks,
linkcolor=darkred,
filecolor=darkgreen,
urlcolor=darkblue,
citecolor=darkblue,
linktocpage=true }

\definecolor{markcolor}{rgb}{.25,0,1}

\definecolor{markcolor2}{rgb}{1,0,0}

\definecolor{markcolor3}{rgb}{0,1,0}

\newcommand{\no}{\noindent}

\def\be{\begin{equation}}
\def\ee{\end{equation}}
\def\ba{\begin{eqnarray}}
\def\ea{\end{eqnarray}}

\def\G{\Gamma}

\def\e{\epsilon}

\def\m{\mu}
\def\n{\nu}
\def\om{\omega}

\def\l{\lambda}
\def\L{\Lambda}
\def\s{\sigma}

\begin{document}
\begin{titlepage}
\begin{center}
\strut\hfill
\vskip 1.3cm


\vskip .5in

{\Large \bf Scattering matrices in the $\mathfrak{sl}(3)$ twisted Yangian}

\vskip 0.5in

{\bf Jean Avan$^{a}$, Anastasia Doikou$^{b,c}$ and Nikos Karaiskos$^{d}$} \vskip 0.2in

{\footnotesize $^{a}$ Laboratoire de Physique Th\'eorique et Mod\'elisation
(CNRS UMR 8089), \\
Universit\'e de Cergy-Pontoise, F-95302 Cergy-Pontoise, France}
\\[4mm]
\noindent
{\footnotesize  $^b$Department of Mathematics, Heriot-Watt University,
\\ EH14 4AS, Edinburgh, United Kingdom}
\\[4mm]
\noindent
{\footnotesize $^{c}$ Department of Computer Engineering \& Informatics, \\
University of Patras, GR-Patras 26500, Greece}
\\[4mm]
\noindent
{\footnotesize  $^d$Institute for Theoretical Physics, Leibniz University
Hannover,\\ Appelstra\ss e 2, 30167 Hannover, Germany}

\vskip .1in


{\footnotesize {\tt E-mail: avan@u-cergy.fr, a.doikou@hw.ac.uk,
nikolaos.karaiskos@itp.uni-hannover.de}}\\

\end{center}

\vskip 1.0in

\centerline{\bf Abstract}

A quantum spin chain with non-conventional boundary conditions is studied.
The distinct nature of these boundary conditions arises from the
conversion of a soliton to an anti-soliton after being reflected
to the boundary, hence the appellation soliton non-preserving
boundary conditions. We focus on the simplest non-trivial case of
this class of models based on the twisted Yangian quadratic algebra.
Our computations are performed through the Bethe ansatz equations in
the thermodynamic limit. We formulate a suitable quantization
condition describing the scattering process and proceed with
explicitly determining the bulk and boundary scattering amplitudes.
The energy and quantum numbers of the low lying excitations are
also derived.

\no

\vfill

\end{titlepage}
\vfill \eject

\section{Introduction}

The description of quantum integrable systems with boundaries dates back to the works of
Cherednik \cite{cherednik} and Sklyanin \cite{sklyanin}. The main object is the
so-called quantum reflection algebra, defined by the quadratic exchange relations
\be
R_{12}\ K_1\ R_{21}\ K_2 = K_2\ R_{12}\ K_1\ R_{21} \, ,
\label{reflection}
\ee
where $R_{12}$ is the bulk quantum $R$-matrix satisfying the Yang-Baxter equation,
$K$ encodes the boundary effects, and the dependence on the spectral parameter is
suppressed throughout the section. Equation (\ref{reflection}) is interpreted as a
supplementary consistency condition between the bulk $S$-matrices and the reflection
matrix $K$ for factorizability of $N$-body amplitudes into 2-body amplitudes
encompassing boundary effects \cite{sklyanin}. In this particular case one interprets
the theory as a description of soliton dynamics with a single bulk collision
2-body $S$-matrix $R_{12}$ and a reflection matrix $K$ preserving the soliton after
reflection, hence the characterization ``soliton-preserving'' boundaries.

The maximal generalization of Eq. (\ref{reflection}) was proposed by Freidel
$\&$ Maillet in \cite{fre-mai}, see also \cite{kulish}. It is parametrized by
three matrices $A,\ B,\ D$
\be
\begin{split}
& \quad A_{12}\ K_1\ B_{12}\ K_2 = K_2\ C_{12}\ K_1\ D_{12} \cr
& A_{12} \ A_{21}={\mathbb I} = D_{12}\ D_{21}, ~~~~~C_{12} =B_{21}.
\label{quadratic}
\end{split}
\ee

When describing the abstract quadratic exchange algebra, $K$ is here interpreted
as a matrix (on auxiliary space 1 or 2) of generators of the quadratic exchange
algebra. It can be systematically constructed from the comodule structure of
(\ref{quadratic}), as ``dressing'' of an initial $K$-scalar solution of
(\ref{quadratic}) by successive left/right ``coproducts'' of pairs $A/C$ or $B/D$.
Within some general assumptions it can be shown reciprocally that all representations,
at least of the reflection algebra (\ref{reflection}), are obtained precisely by
the dressing of a scalar $K$-matrix by bulk quantities obeying a Yang-Baxter
type equation \cite{more}.

We shall focus here on another particular case of (\ref{quadratic}), when the
initial reflection on the boundary exhibits a soliton non-preserving behavior e.g.
when the reflection converts a soliton into an antisoliton
(see e.g. \cite{Doikou:2000yw, Arnaudon:2004sd} and references therein). In this
framework one is lead to identify $A_{12} =R_{12},\ D_{12} =R_{21}$ and $B_{12} = \bar R_{12}$
where physically $R_{12}$ corresponds to the soliton$-$soliton collision matrix,
whereas $\bar R_{12}$ corresponds to the soliton$-$anti-soliton collision matrix.
We get then the following structure (see also \cite{twisted})
\be
R_{12}\ K_1\ \bar R_{2 1}\ K_2 = K_2\ \bar R_{1 2}\ K_1 \ R_{21} \,.
\label{twist}
\ee

A suitable double-row monodromy matrix is then defined as alternated coproducts as
commented before in the general case
\be
{\mathbb T} = \ldots R_{02}\ \bar R_{01}\ K^-_0\ R_{10}\ \bar R_{20} \ldots \, ,
\label{mod_mon}
\ee
and the relevant spin chain Hamiltonians are now obtained from the quantum trace formula:
\be
\tau = Tr\Big \{ K^+\ {\mathbb T} \Big \} \, .
\ee
Assuming $R$ possesses the regularity property
\be
R_{12}(\l \to 0) \propto \mathcal{P}_{12} \, ,
\ee
with $\mathcal{P}$ being the permutation operator and $\l$ denoting the spectral
parameter, the Hamiltonian
\be
H_1 \propto {d \over d\l} (\ln \tau(\l))\Big|_{\l=0} \, ,
\ee
yields a local spin chain interaction with boundary terms. It has the following explicit form  (for more details see \cite{Doikou:2000yw}, \cite{Arnaudon:2004sd}):
\begin{eqnarray}
  {\cal H} &\propto &\sum_{j=1}^L \bar R'_{2j-1\ 2j}\ \bar R_{2j-1\
  2j} + \sum_{j=1}^{L -1}\bar R_{2j+1\ 2j+2}\ \check R'_{2j\ 2j+2}\
  \bar R_{2j+1\ 2j+2} \nonumber \\
  &+& \sum_{j=1}^{L -1}\bar R_{2j+1\ 2j+2}\ \bar R_{2j-1\ 2j}\ \bar
  R'_{2j-1\ 2j+2}\ \bar R_{2j-1\ 2j+2}\ \bar R_{2j-1\ 2j}\ \bar R_{2j+1\
  2j+2} \nonumber \\
  &+& \sum_{j=1}^{L -1}\bar R_{2j+1\ 2j+2}\ \bar R_{2j-1\ 2j}\ \bar R_{2j-1\
  2j+2}\ \check R'_{2j-1\ 2j+1}\ \bar R_{2j-1\ 2j+2}\ \bar R_{2j-1\ 2j}\
  \bar R_{2j+1\ 2j+2} \nonumber \\
  &+& Tr_{0}\check R'_{0\ 2L}\ \bar R_{2L-1\ 2L}\ {\cal P}_{0\ 2L-1}\ \bar
  R_{0\ 2L-1}\ \bar R_{2L-1\ 2L} + \bar R_{12}\ \check R'_{12}\ \bar R_{12},
  \label{ham2}
\end{eqnarray}
The prime denotes the derivative with respect to the spectral parameter and $\check R = {\cal P}\ R$. Note that this type of unconventional boundary conditions
in the quantum spin chain framework were first studied in \cite{Doikou:2000yw} and later generalized in \cite{Arnaudon:2004sd}. These boundary conditions were originally known, albeit in a classical framework, in the context of affine Toda field theories \cite{corrigan}. It is worth pointing out that the implementation of these boundary conditions, based on the twisted Yangian, in the quantum spin chain frame provides a resolution of a long lasting misunderstanding regarding the various types of boundary conditions in integrable classical field theories vs. integrable lattice models. More precisely, until the full study of all possible conditions in field theories \cite{doikou-affine} and quantum spin chains \cite{Doikou:2000yw} only boundary conditions associated to the reflection algebra were know in the spin chain context, whereas in affine Toda field theories only boundary conditions associated to the classical twisted Yangian were known.

Here we propose for the first time to study such systems in the thermodynamic limit aiming at this time at computing the bulk and boundary scattering amplitudes after implementing a novel quantization condition related to the particular models.
We shall here concentrate
on the special case associated to $\mathfrak{sl}(3)$. In addition, we consider a
case where the conjugate $R$-matrix $\bar R$ is obtained from $R$ by
\be
\bar R_{12} = V_1\ R_{12}^{t_2}\ V_1 \, .
\ee
The relevant algebraic structure (\ref{twist}) is now identified as a twisted
Yangian (rational $R$-matrix) or twisted quantum Yangian (trigonometric case.)

Remark that a natural construction of representations of the twisted Yangian consists
in starting from the bulk monodromy matrix $T$ obeying the fundamental quadratic
relation \cite{fad}
\be
R_{12}\ T_1\ T_2 = T_2\ T_1\ R_{12} \, ,
\ee
and define the ``folded" or twisted generic $K$ matrix as:
\be
{\mathbb K}(\l) = T(\l)\, K(\l)\ T^t(-\l +\kappa) \, ,
\ee
where $K$ is $c$-number solution of the twisted Yangian equation, $\kappa$ is a constant
associated
with the Lie algebra of the chosen $R$-matrix, and $^t$ denotes the transposition
taken on the auxiliary space only. This natural ``folding" structure is also seen
in the formula (\ref{mod_mon}), and will have consequences on the structure of
the vacuum, the eigenvectors as well as the exact symmetry of the corresponding
integrable system.

This article is organized as follows. In the next Section we focus on the $\frak{sl}(3)$
twisted Yangian model and study its thermodynamic limit. We compute the energy
of an excitation and study the quantum numbers in order to ensure the validity
of our results. Section 3 contains the main results of our work. The key result is the formulation of
a quantization condition for twisted Yangian spin chains; we then prove the
factorization of the bulk scattering amplitude and explicitly compute the
boundary scattering amplitude. Note that the results of this Section are completely
new. We conclude with a short discussion.

\section{Twisted Yangian: Bethe ansatz and thermodynamics}

The twisted Yangian algebra associated to the so-called soliton non-preserving
boundary conditions was first studied in the context of integrable lattice models
via the Bethe ansatz formulation in \cite{Doikou:2000yw}, whereas
generalizations were investigated in \cite{Arnaudon:2004sd}.
It was
shown in \cite{Doikou:2000yw} that the Bethe ansatz equations (BAE) of
the model are given as
\be
e_1(\l_i)^{L} \, e_{-\frac{1}{2}}(\l_i) = -
\prod_{j=1}^M e_2(\l_i-\l_j) \, e_2(\l_i+\l_j)
\, e_{-1}(\l_i-\l_j) \, e_{-1}(\l_i+\l_j) \, ,
\label{BAE}
\ee
where we define
\be
e_n(\l) = \frac{\l + \frac{in}{2}}{\l - \frac{in}{2}} \, .
\ee
These BAE are similar to those of the $\mathfrak{osp}(1|2)$ case
\cite{Arnaudon:2003zw}, up to an extra
boundary contribution. In fact, the case in study is the
first occurence of a more general correspondence between $\mathfrak{sl}(2n+1)$
chains with twisted Yangian boundary conditions and $\mathfrak{osp}(1|2n)$
open spin chains with certain boundary conditions. This correspondence
was already studied in \cite{Arnaudon:2003zw}, and is currently
under investigation \cite{ADNK} from the Bethe ansatz point of view.

In the usual $\mathfrak{sl}(2n+1)$ Yangian case the ground state of the system
consists of $2n$ filled Dirac seas. On the contrary, in the twisted
Yangian case, this number is halved, due to the ``folding''. The
bulk contribution is essentially the same as that of a spin chain
with $\mathfrak{osp}(1|2n)$ symmetry, hence the intriguing correspondence
mentioned above.

The $\mathfrak{sl}(3)$ twisted Yangian quantum spin chain in particular
has only one filled Dirac sea as its ground state. A hole in the Dirac sea represents
an excitation in the system and incorporates both the fundamental 3
and its conjugate $\bar 3$ representation of $\mathfrak{sl}(3)$, i.e. both a soliton
and an anti-soliton are present in an excitation. The thermodynamic
limit is performed according to the rule
\be
\frac{1}{L} \sum_{j=1}^M f(\l_j) \to
\int_0^{\infty} d\m \, \s(\m) \, f(\m) -
\frac{1}{L} \sum_{j=1}^\n f(\tilde{\l}_j)
- \frac{1}{2L} f(0)\, ,
\ee
for $\n$ holes in the Dirac sea with rapidities $\tilde{\l}_j$, and the
last term is the halved contribution at $0^+$ due to the boundaries.
Defining also
\be
a_n(\l) = \frac{1}{2\pi}\frac{d}{d\l}\ln e_n(\l) \, ,
\ee
the density of the Bethe roots as computed from the BAE (\ref{BAE})
is given by (see also \cite{Yang:1967bm, Korepin:1979hg, Andrei:1983cb})
\be
\begin{split}
\s(\l) & =  a_1(\l) -
\int_{-\infty}^\infty d\m \, \s(\m)  \Big ( a_2(\l-\m)
- a_{1}(\l-\m) \Big ) \cr
& + \frac{1}{L} \sum_{j=1}^\n \Big(
a_2(\l - \tilde{\l}_j) + a_2 (\l + \tilde{\l}_j)
- a_{1}(\l - \tilde{\l}_j) - a_{1}(\l+\tilde{\l}_j)
\Big) \cr
& + \frac{1}{L}\big(a_2(\lambda) - a_{1}(\l) - a_{\frac{1}{2}}(\l) \big) \, .
\end{split}
\ee
Taking the Fourier transform\footnote{
The following Fourier conventions are used
\be
\hat{f}(\om) = \int_{-\infty}^\infty dx \, f(x) \, e^{i\om x} \, ,
\qquad
f(x) = \frac{1}{2\pi} \int_{-\infty}^\infty d\om \, \hat{f}(\om)
\, e^{-i\om x} \, .
\ee
} of the latter expression leads to
\be
\hat{\mathcal{K}}(\om) \, \hat{\s}(\om) =
\hat{a}_1(\om)
+ \frac{1}{L}
\left[
\hat{a}_2(\om) - \hat{a}_1(\om) - \hat{a}_{\frac{1}{2}}(\om)
+ \big(\hat{a}_2(\om) - \hat{a}_1(\om)\big)
\sum_{j=1}^\n
(e^{i\om\tilde{\l}_j} + e^{-i\om\tilde{\l}_j})
\right]
\, ,
\ee
where we have defined the kernel
\be
\hat{\mathcal{K}}(\om) \equiv 1 +\hat{a}_2(\om) - \hat{a}_1(\om)
= e^{-\frac{|\om|}{2}} \frac{\cosh \frac{3\om}{4}}{\cosh\frac{\om}{4}} \, ,
\label{kernel} \qquad \mbox{and} \qquad\hat{a}_n(\om) =  e^{-\frac{n |\omega|}{2}}.
\ee
After some simplifications, the density is written compactly as
\be
\hat{\s}(\om) = \hat{\sigma}^{(0)}(\om) + \frac{1}{L}
\left[
\hat{r}_1(\om) + \hat{r}_2(\om)
\sum_{j=1}^\nu
(e^{i\om\tilde{\l}_j} + e^{-i\om\tilde{\l}_j})
\right]
\, ,
\ee
with
\be
\hat{\sigma}^{(0)}(\om) \equiv
\frac{\cosh\frac{\om}{4}}{\cosh\frac{3\om}{4}} \, ,
\quad
\hat{r}_1(\om) \equiv
\big(
e^{-\frac{|\om|}{2}} - e^{\frac{|\om|}{4}} - 1
\big)
\hat{\sigma}^{(0)}(\om) \, ,
\quad
\hat{r}_2(\om) \equiv
\big(e^{-\frac{|\om|}{2}} - 1
\big)
\hat{\sigma}^{(0)}(\om)
\, .
\ee
The first term of the density turns out to be the energy of the ground state,
as will be transparent later in the
text, and coincides with that of
the $\mathfrak{osp}(1|2)$ spin chain, while the terms $\hat{r}_1$ and $\hat{r}_2$
correspond to boundary and bulk scattering contributions. In coordinate space we may write
\be
\s(\l) = \sigma^{(0)}(\l) + \frac{1}{L}
\left[
r_1(\l)
+ \sum_{j=1}^\nu
\big(
r_2(\l-\tilde{\l}_j) + r_2(\l + \tilde{\l}_j)
\big)
\right]
\, . \label{density1}
\ee
The latter expression will be used subsequently for the computation of
the energy and quantum numbers of the low lying excitations as well as
the computation of the bulk and boundary scattering amplitudes.

\subsection{The energy}

It will be instructive to derive the energy of the excitations as well as the
relevant quantum numbers associated to these states.
We first derive the energy eigenvalue directly
from the algebraic Bethe Ansatz, by focusing on one excitation-hole in
the system.
This computation also serves as an extra validity check of our
computations regarding the ground state and the low-lying excitations.
Recall that the eigenvalues of the transfer matrix are given by
\cite{Doikou:2000yw}
\be
\L^{(M)}(\l) = \big(a(\l)\bar{b}(\l)\big)^{L}
\frac{\bar{a}(2\l)}{\bar{b}(2\l)} A_1(\l)
+ \big(b(\l)\bar{b}(\l)\big)^{L}A_2(\l)
+\big(\bar{a}(\l)b(\l)\big)^{L}
\frac{\bar{a}(2\l+2i)}{\bar{b}(2\l)} A_3(\l) \, ,
\ee
where
\be
\begin{split}
&a(\l) = \l + i \, ,\qquad b(\l) = \l \, , \qquad c(\l) = i \, , \cr
& \qquad A_1(\l) = \prod_{j=1}^M \frac{\l+\m_j - \frac{i}{2}}{\l + \m_j + \frac{i}{2}}
\frac{\l - \m_j - \frac{i}{2}}{\l - \m_j + \frac{i}{2}} \, ,
\end{split}
\label{abc_def}
\ee
$\{\m_j\}$ is the set of Bethe roots and $\bar{f}(\l) = f(-\l - \frac{3i}{2})$.
The terms containing $A_2(\l), A_3(\l)$, as well as their derivatives
vanish for $\l=0$ and hence are not needed here. The exact expressions for
$A_2(\l), A_3(\l)$ can be found in \cite{Doikou:2000yw}
The first derivative of the eigenvalues with respect to the spectral parameter
yields the energy of the system
\be
\begin{split}
E(\{\m_j\}) & \propto \frac{d}{d\l}\L(\l, \{\m_j\}) \Big|_{\l=0}  \cr
& =
\frac{d}{d\l}\left[\big(a(\l) \bar{b}(\l)\big)^{L}\right]_{\l=0}  A_1(0)
+ \big(a(0) \bar{b}(0)\big)^L A_1'(\{\m_j\}) \, ,
\end{split}
\ee
with
\be
A_1'(\{\m_j\}) = \frac{d}{d\l}
A_1(\l)
\Big|_{\l=0} = \sum_{j=1}^M \frac{2i}{\m_j^2 + \frac{1}{4}}
= -4\pi \sum_{j=1}^M a_1(\m_j)\, .
\ee
Since $A_1(0)=1$, the first term contributing to the energy is independent
of the Bethe roots and thus corresponds to a simple energy shift. We may
then conclude that
\be
E(\{ \m_j \}) = - \sum_{j=1}^M a_1(\m_j) \, .
\ee
In the thermodynamic limit and in the case of one hole present in the system,
the above relation takes the form
\be
\e(\tilde{\l}_1) = - \int_{0}^\infty d\m \, a_1(\m) \, \s(\m)
+ \frac{1}{L} a_1(\tilde{\l}_1) - \frac{1}{2L} a_1(0) \, .
\ee
For our purpose here the boundary contribution is irrelevant, since it
only contributes to the ground state. Gathering the Fourier transformed
$\frac{1}{L}$ contributions containing the rapidity of the excitation,
$\tilde{\l}_1$, one concludes that
\be
\hat{\e}(\om) =
\frac{\hat{a}_1(\om)}{1 + \hat{a}_2(\om) - \hat{a}_1(\om)}
= \hat{\s}^{(0)}(\om) \, ,
\ee
which as expected coincides with ground state density --up to boundary
contributions. This is a key
point for the computation of scattering amplitudes via the suitable
quantization condition, which will be formulated later on in the text.

\subsection{Quantum numbers \& symmetry}

As was discussed in detail in \cite{Doikou:2000yw, Arnaudon:2004sd}
from the study of the asymptotics
of the transfer matrix one can extract the total spin:
\be
{\mathrm S} = \sum_{j=1}^L S_j^z  = L - M - \frac{1}{2} \, , \qquad
S^z =
\left(
\begin{array}{ccc}
 1 & 0 & 0 \cr
 0 & 0 & 0 \cr
 0 & 0 & -1
\end{array}
\right) \, .
\ee
Through the thermodynamic limit computations, for a state with $\n$
holes
we have
\be
M = \sum_{j=1}^M 1 = L \int_0^\infty \s(\l)  \, d\l -
\underbrace{\n}_{\textrm{holes}}
- \underbrace{\frac{1}{2}}_{\substack{\textrm{boundary} \\ \textrm{effect}}}
= L - \n -\frac{1}{2} \, .
\ee
For a state with one hole, the spin of this state is indeed correctly
computed to be ${\mathrm S}=1$. Considering now the state with two
holes, the co-product is computed
\be
\mathrm{S} = S^z \otimes \mathbb{I} + \mathbb{I} \otimes S^z =
\left(
\begin{array}{ccc|ccc|ccc}
 2 & & & & & & & & \cr
  & 1 & & & & & & & \cr
  & & 0 & & & & & & \cr
 \hline
  & & & 1 & & & & \cr
  & & & & 0& & & \cr
  & & & & & -1 & & \cr
 \hline
  & & & & & &  0 & &  \cr
  & & & & & & & -1 & \cr
  & & & & & & & & - 2
\end{array}
\right) \, .
\label{coprod}
\ee
Recall that the $\mathfrak{sl}(3)$ invariant $R$-matrix is given by \cite{Yang:1967bm}
\be
R(\l)  = a(\l) \sum_{i=1}^3 e_{ii} \otimes e_{ii}
+ b(\l) \sum_{i \neq j}^3 e_{ii} \otimes e_{jj}
+ c(\l) \sum_{i \neq j}^3 e_{ij} \otimes e_{ji} \, ,
\ee
where $a,b,c$ were defined in Eq. (\ref{abc_def}), while its conjugate is
defined as
\be
\bar{R}_{12}(\l)  =
V_1 \, R_{12}^{t_2}(-\l - \tfrac{3i}{2}) \,  V_1 \, , \qquad
V =
\left(
\begin{array}{ccc}
 &  & 1 \cr
 & 1 &  \cr
1 &  &
\end{array}
\right) \, .
\ee
%
We are interested in identifying the common eigenvectors of the
co-product state (\ref{coprod}) and the product $R(\l)\bar{R}(\l)$. Let us
then introduce the following basis of the real vector space
$\mathbb{R}^3$
\be
|\textrm{+}1\rangle = \left(
\begin{array}{c}
 1 \cr
 0 \cr
 0
\end{array}
\right) \, , \qquad
|0\rangle = \left(
\begin{array}{c}
 0 \cr
 1 \cr
 0
\end{array}
\right) \, , \qquad
|\textrm{--}1\rangle = \left(
\begin{array}{c}
 0 \cr
 0 \cr
 1
\end{array}
\right) \, .
\ee
We identify the following common eigenvectors and associated
eigenvalues:
\be
\begin{tabular}{|c|c|c|}
\hline
Spin & Eigenvector & Eigenvalue \cr
\hline \hline
+2 & $|\textrm{+}1\rangle \otimes |\textrm{+}1\rangle$
& $a(\l) \, \bar{b}(\l)$
\cr
+1 & $|\textrm{+}1\rangle \otimes |0\rangle
+ |0\rangle \otimes |\textrm{+}1\rangle $
& $\bar{b}(\l) \big(b(\l)+c(\l)\big)$\cr
0 & $|\textrm{+}1\rangle \otimes |\textrm{--}1\rangle
+ |0\rangle \otimes |0\rangle
+ |\textrm{--}1\rangle \otimes |\textrm{+}1\rangle $
& $a(\l)\big(\bar{a}(\l) + 2c(\l)\big)$ \cr
$-1$ & $|\textrm{--}1\rangle \otimes |0\rangle
+ |0\rangle \otimes |\textrm{--}1\rangle $
& $\bar{b}\big(\l) (b(\l)+c(\l)\big)$\cr
$-2$ & $|\textrm{--} 1\rangle \otimes |\textrm{--}1\rangle$
&  $a(\l) \, \bar{b}(\l)$ \cr
\hline
\end{tabular}
\ee

\section{Scattering amplitudes}

The main aim in this section is the study of the bulk and boundary
scattering for the $\mathfrak{sl}(3)$ twisted Yangian model.
To achieve this we shall basically employ the results of the previous
section together with a suitable quantization condition (see also \cite{Korepin:1979hg, Andrei:1983cb, Grisaru:1994ru}).
Thus before we proceed with the computation of exact $S$-matrices via
the twisted Yangian BAE it will be important to formulate the associated quantization condition, which describes the bulk
and boundary scattering in the particular algebraic setting.

\subsection{Quantization condition}

Here we shall derive the suitable quantization case associated to the
soliton non-preserving scattering. This is in fact one of the key points
in the present article, and it is also a starting point for the investigation
of the bulk and boundary scattering.

It is constructive to graphically depict the scattering matrices in order to
fully comprehend the quantization condition. A soliton will be represented by
a solid line and an anti-soliton by a dashed one.
Let $S$ denote the soliton$-$soliton (or anti-soliton$-$anti-soliton)
and $\bar{S}$ denote the soliton$-$anti-soliton scattering
respectively. They are depicted as
\begin{picture}(0,0)(300,75)

\put(0,25){\line(1,0){50}}
\put(25,50){\line(0,-1){50}}

\put(65,22){$\simeq$}
\put(65,45){$S$}

\put(-12,25){\multiput(100,0)(7,0){8}{\line(1,0){3}}}
\put(14,50){\multiput(100,0)(0,-7){8}{\line(0,-1){2}}}

\put(170,22){and}

\put(220,25){\line(1,0){50}}
\put(145,50){\multiput(100,0)(0,-7){8}{\line(0,-1){3}}}

\put(285,22){$\simeq$}
\put(285,45){$\bar{S}$}

\put(210,25){\multiput(100,0)(7,0){8}{\line(1,0){3}}}
\put(336,50){\line(0,-1){50}}

\end{picture}
\vskip 3cm

Before we discuss the quantization condition associated to the twisted
Yangian let us first recall the quantization condition for
the usual reflection case \cite{Grisaru:1994ru}. The double-row
transfer matrix consists of two products of the bulk $S$-matrix,
intertwined with the reflection matrices $K^{\pm}$. A graphical
illustration of such a model is given as

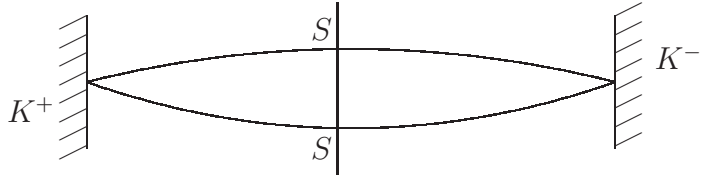
\begin{picture}(0,0)(-125,60)

\put(0,50){\line(0,-1){50}}
\multiput(0,50)(0,-7){8}{\line(-2,-1){10}}

\qbezier(0,25)(100,-10)(200,25)
\qbezier(0,25)(100,50)(200,25)

\put(200,50){\line(0,-1){50}}
\multiput(200,50)(0,-7){8}{\line(2,1){10}}

\put(95,55){\line(0,-1){65}}

\put(-30,10){$K^+$}
\put(85,41){$S$}
\put(85,-4){$S$}
\put(215,30){$K^-$}

\end{picture}
\vskip 2.7cm

\no One imposes an isomonodromy condition on the state of two holes as:
\be
\Big(e^{2i\mathcal{P}L} \,
\mathbb{S}(\tilde \l_1, \tilde \l_2) - 1 \Big)
| \tilde \l_1, \tilde \l_2\rangle = 0 \, ,
\label{mom_quant}
\ee
where the global scattering amplitude $\mathbb{S}$ is given as:
\be
\mathbb{S}(\l_1,\l_2) \equiv
K^+(\l_1) \, S(\l_1-\l_2) \,
K^-(\l_1) \, S(\l_1+\l_2) \, .
\label{quantiz_cond}
\ee

We come now to our main  objective which is the derivation of a generalized
quantization condition regarding the soliton non-preserving equation. Recall
that the transfer matrix of the model consists of alternated coproducts, as
mentioned in the introduction. A graphical illustration of a model with
twisted Yangian boundary conditions will have the following form then

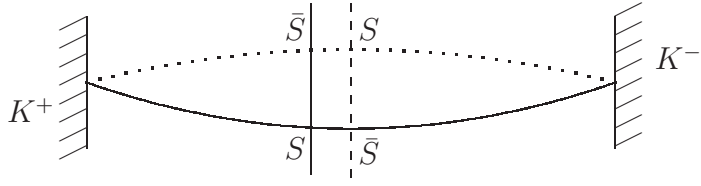
\begin{picture}(0,0)(-125,60)

\put(0,50){\line(0,-1){50}}
\multiput(0,50)(0,-7){8}{\line(-2,-1){10}}

\qbezier(0,25)(100,-10)(200,25)

\linethickness{0.3mm}

\qbezier[40](0,25)(100,50)(200,25)

\linethickness{0.1mm}

\put(200,50){\line(0,-1){50}}
\multiput(200,50)(0,-7){8}{\line(2,1){10}}

\put(85,55){\line(0,-1){65}}
\multiput(100,55)(0,-7){10}{\line(0,-1){3}}

\put(-30,10){$K^+$}
\put(75,41){$\bar{S}$}
\put(103,41){$S$}
\put(75,-4){$S$}
\put(103,-6){$\bar{S}$}
\put(215,30){$K^-$}

\end{picture}
\vskip 2.7cm

\noindent from which the momentum quantization condition follows directly again as an isomonodromy condition
\be
\Big(e^{i\mathcal{P}L} \,
\mathbb{S}(\tilde \l_1, \tilde \l_2) - 1 \Big)
| \tilde \l_1, \tilde \l_2\rangle = 0 \, ,
\label{mom_quant}
\ee
with the manifest factorization
\be
\mathbb{S}(\l_1,\l_2) \equiv
k^+(\l_1) \, S(\l_1-\l_2) \, \bar{S}(\l_1-\l_2) \,
k^-(\l_1) \, S(\l_1+\l_2) \, \bar{S}(\l_1+\l_2) \, ,
\label{quantiz_cond}
\ee
and $L$ being the length of the chain. Note that the phase in the
exponential factor is just $L$ instead of the usual $2L$, because we deal here with
``folding'' and not reflection, as opposed to the usual open boundary
conditions. The ``particle" --merging of 3 and $\bar 3$-- now propagates in both directions simultaneously, hence now over a distance $L$. This factorization is expected to emerge naturally at the
thermodynamic limit. Indeed, we show below that the bulk scattering
amplitudes factorize appropriately, which confirms the quantization
condition as formulated in (\ref{mom_quant}).

From now on we consider two excitations (holes), so that $\nu=2$.
Recall that the momentum and energy are related through
\be
\e(\l) = \frac{1}{2\pi} \frac{dp}{d\l} \,.
\ee
Combining the momentum quantization condition (\ref{mom_quant})
with the above expression, and
taking into account that
\be
L \int_0^{\tilde{\l}_1} d\l \, \s(\l) \in \mathbb{Z} \, ,
\ee
we find that the scattering matrix phase, $\mathbb{S} = \exp(i\Phi)$,
is computed through
\be
\Phi = 2\pi\int_{0}^{\tilde{\l}_1} d\l \,
\left[
r_1(\l)
+ \sum_{j=1}^2
\big(
r_2(\l-\tilde{\l}_j) + r_2(\l + \tilde{\l}_j)
\big)
\right] \, ,
\ee
or passing to momentum space to perform the computations
\be
\begin{split}
\Phi =
- \int_{-\infty}^\infty \frac{d\om}{\om}
\Big(
e^{-i\om \tilde{\l}_1} \, \hat{r}_1(\om) +
e^{-2i\om \tilde{\l}_1} \, \hat{r}_2(\om)
\Big)
- 2 \int_{-\infty}^\infty \frac{d\om}{\om}
e^{-i\om \tilde{\l}_1}
\cos(\om\tilde{\l}_2)
\, \hat{r}_2(\om)
\, .
\end{split}
\ee
The first integral provides the boundary contribution and the second
one the bulk scattering. Recalling the quantization condition, one
obtains
\be
\begin{split}
 k^+(\l)\, k^-(\l) = &
 \exp\left[
 - \int_{-\infty}^\infty \frac{d\om}{\om}
\Big(
e^{-i\om \l} \, \hat{r}_1(\om) +
e^{-2i\om \l} \, \hat{r}_2(\om)
\Big)
 \right] \cr
 \cal{S}(\l)  = &
 \exp\left[
- \int_{-\infty}^\infty \frac{d\om}{\om}
\, \hat{r}_2(\om) \, e^{-i\om \l}
\right] \, .
\end{split}
\label{bulk_bound}
\ee
where we recall that $k^{\pm}$ correspond to the left and right boundary
scattering amplitudes, and ${\cal S}$ is the bulk scattering amplitude. We have considered here for simplicity $K^{\pm}\propto {\mathbb I}$, so identifying the scattering amplitude $k^{\pm}$ suffices, i.e. $K^{\pm}(\l) = k^{\pm}(\l){\mathbb I}$. As
will be clear subsequently the bulk scattering factorizes into
soliton$-$soliton amplitudes times the soliton$-$anti-soliton amplitude.

\subsection{Bulk scattering amplitude: factorization}

Let us first focus on the bulk scattering and verify that the scattering
factorizes into the two amplitudes mentioned above.
After some algebra, it can be shown that the integrand in the
bulk scattering amplitude appearing in Eq. (\ref{bulk_bound}) is
given by
\be
\hat{r}_{\cal S}(\om) =
\hat{r}_2(\om) =
\frac{(e^{\frac{\om}{2}} - e^{-\frac{\om}{2}})
(e^{-\om} - e^{\frac{\om}{2}})}{2\sinh\frac{3\om}{2}} \, .
\label{twistamp}
\ee
This expression should be compared with the expressions computed in
the Yangian $\mathfrak{sl}(3)$ model. More specifically, the soliton$-$soliton
and soliton$-$anti-soliton amplitudes in that model are
given by the following expressions
\be
\begin{split}
 & \hat{r}_S(\om) = \hat{a}_2(\om) \hat{\mathcal{R}}_{11}(\om)
 - \hat{a}_1(\om) \hat{\mathcal{R}}_{12}(\om) =
 \frac{e^{-\frac{\om}{2}}-e^{-\frac{3\om}{2}}}{2\sinh\frac{3\om}{2}} \, , \cr
 & \hat{r}_{\bar{S}}(\om) = \hat{a}_2(\om) \hat{\mathcal{R}}_{12}(\om)
 - \hat{a}_1(\om) \hat{\mathcal{R}}_{11}(\om) =
 \frac{1 - e^\om}{2\sinh\frac{3\om}{2}} \, ,
\end{split}
\label{1nm1amp}
\ee
where $\hat{\mathcal{R}}_{ij}(\om)$ denotes the inverse of the kernel
for the bulk $\mathfrak{sl}(3)$ scattering \cite{Suth1975}
\be
\hat{\mathcal{R}}_{ij}(\om) = e^{\frac{|\om|}{2}}
\frac{\sinh\big(\min(i,j)\frac{|\om|}{2}\big)
\sinh \big[ \big(3-\max(i,j)\big) \big] \frac{|\om|}{2}}
{\sinh\frac{|\om|}{2} \sinh \frac{3|\om|}{2}} \, .
\ee
A quick inspection of relations (\ref{twistamp}) and (\ref{1nm1amp}) reveals that
\be
\hat{r}_{\cal S}(\om) = \hat{r}_S(\om) + \hat{r}_{\bar{S}}(\om) \,
\Rightarrow \ {\cal S}(\lambda) = S(\lambda)\ \bar S(\lambda) \label{factor} \, ,
\ee
where we define:
\be
\mathcal{X}(\l) =
\exp \left[ -\int_{-\infty}^{\infty} {d\omega \over
\omega} \, \hat{r}_{\mathcal{X}}(\omega) \, e^{-i\omega \lambda} \right] \, ,
\qquad \mathcal{X} \in \{ \mathcal{S}, S, \bar{S}\} \, .
\ee
Relation (\ref{factor}) expresses the expected factorization of the bulk
amplitude into
two separate ones, the soliton$-$soliton and soliton$-$anti-soliton amplitude,
(see \cite{doikou-nepo} and references therein), which correspond to
$S(\l)$ and $\bar{S}(\l)$. This fact confirms the validity of the form of the
quantization condition as formulated in (\ref{quantiz_cond}).

\subsection{Boundary scattering amplitude}

Let us now come to the study of the boundary scattering.
Recalling the first relation of (\ref{bulk_bound}),
we denote the boundary amplitude as
\be
k^+(\l) \, k^-(\l)
=
 \exp\left[
 - \int_{-\infty}^\infty \frac{d\om}{\om}
\Big(
e^{-i\om \l} \, \hat{r}_1(\om) +
e^{-2i\om \l} \, \hat{r}_2(\om)
\Big)
 \right]
= \exp\Big[
\mathcal{A}_1 + \mathcal{A}_2
\Big] \, .
\ee
It is convenient here as in the bulk
case to express the boundary scattering amplitudes in terms
of $\Gamma$-functions. For that purpose we use the identity
\be
\frac{1}{2} \int_0^\infty \frac{d\om}{\om}
\frac{e^{-\frac{\m\om}{2}}}{\cosh \frac{\om}{2}}
= \ln \frac{\G(\frac{\m+1}{4})}{\G(\frac{\m+3}{4})} \, ,
\label{ident}
\ee
and we therefore express the amplitude in the form
%
%
\be
{\cal A}_1 = - \frac{1}{2} \int_0^\infty \frac{d\om}{\om}
\frac{e^{-i\om \l} \, \big( e^{-\frac{3\om}{4}} - e^{\frac{\om}{2}}
 - e^{\frac{\om}{4}} -1
\big)}{\cosh\frac{3\om}{4}}
+ \int_0^\infty \big(\l \to - \l\big) \, .
\ee
Using the identity (\ref{ident}) as well as
\be
\G(x)\ \G(1-x) = {\pi \over \sin(\pi x)} \, ,
\ee
we compute the boundary contribution $\mathcal{A}_1$
\be
\begin{split}
S^{(1)} = \exp({\cal A}_1)
& = {\tan {\pi\over 3}(i\lambda -1) \over \tan {\pi\over 3}(i\lambda +1)}
\frac{\G(\frac{i\l}{3} + \frac{1}{12})}{\G(\frac{i\l}{3} + {7\over 12})}
 \frac{\G(\frac{i\l}{3} + \frac{1}{4})}{\G(\frac{i\l}{3} + \frac{3}{4})}
\frac{\G(\frac{i\l}{3} + \frac{10}{12})}
{\G(\frac{i\l}{3} + \frac{4}{12})}
 \frac{\G(\frac{i\l}{3} + 1)}
{\G(\frac{i\l}{3} + \frac{1}{2})} \cr
& \qquad \times \frac{\G(-\frac{i\l}{3} + \frac{7}{12})}{\G(-\frac{i\l}{3} + {1\over
12})}
 \frac{\G(-\frac{i\l}{3} + \frac{3}{4})}{\G(-\frac{i\l}{3} + \frac{1}{4})}
\frac{\G(-\frac{i\l}{3} + \frac{1}{2})}
{\G(-\frac{i\l}{3} + 1)}
 \frac{\G(-\frac{i\l}{3} + \frac{4}{12})}
{\G(-\frac{i\l}{3} + \frac{10}{12})} \, .
\end{split}
\ee
Let us also compute the other term associated to the boundary scattering
amplitude:
%
%
\be
{\cal A}_2 =
 - \frac{1}{2} \int_0^\infty \frac{d\om}{\om}
e^{-2i\om \l} \frac{\big( e^{-\frac{3\om}{4}} -
e^{\frac{\om}{4}}\big)}{\cosh\frac{3\om}{4}} +
\int_0^\infty \big(\l \to - \l\big) \, ,
\ee
Using the identity (\ref{ident}) together with the duplication formula
for the $\Gamma$-function
\be
\G(x)\ \G(x+{1\over 2}) = 2^{-2x+1} \sqrt{\pi} \G(2x) \, ,
\ee
we obtain
\be
\begin{split}
S^{(2)} = \exp({\cal A}_2)& =  \frac{\G(\frac{i\l}{3} + \frac{1}{12})}
{\G(\frac{i\l}{3} + \frac{4}{12})}
 \frac{\G(\frac{i\l}{3} + \frac{7}{12})}{\G(\frac{i\l}{3} + \frac{10}{12})}
 \frac{\G(\frac{i\l}{3} + \frac{1}{2})}{\G(\frac{i\l}{3} + \frac{1}{4})}
 \frac{\G(\frac{i\l}{3} + 1)}{\G(\frac{i\l}{3} + \frac{3}{4})}\cr
 & \qquad \times \frac {\G(-\frac{i\l}{3} + \frac{4}{12})}{\G(-\frac{i\l}{3} + \frac{1}
 {12})}\frac{\G(-\frac{i\l}{3} +  \frac{10}{12})}{\G(-\frac{i\l}{3} + \frac{7}
 {12})}\frac{\G(-\frac{i\l}{3} + \frac{3}{4})}{\G(-\frac{2i\l}{3} + 1)}
 \frac{\G(-\frac{i\l}{3} + \frac{1}{4})}{\G(-\frac{2i\l}{3} + {1\over 2})}\, .
\end{split}
\ee
Finally, the total boundary amplitude associated to the left and right
boundary scattering is given as:
\be
\begin{split}
k^+(\lambda) k^-(\lambda)= S^{(1)} S^{(2)} & =  {\tan {\pi\over 3}(i
\lambda -1) \over \tan {\pi\over 3}(i\lambda +1)}\Big ( \frac{\G(\frac{i\l}
{3} + \frac{1}{12})}{\G(\frac{i\l}{3} + \frac{3}{4})}\frac{\G(\frac{i\l}{3} + 1)}
{\G(\frac{i\l}{3} + \frac{1}{3})} \Big )^2 \cr
& \qquad \times \Big ( \frac{\G(-\frac{i\l}{3} + \frac{3}{4})}{\G(-\frac{i\l}{3} +
\frac{1}{12})}\frac{\G(-\frac{i\l}{3} + \frac{1}{3})}{\G(-\frac{i\l}{3} +1)}
\Big )^2.
\end{split}
\ee
This concludes our derivation of the boundary scattering
amplitude. Notice that since we have chosen the simplest reflection matrices
$K^{\pm} \propto {\mathbb I}$, one only needs to compute
the overall physical factor (amplitude) for the left and right boundary
scattering.

\section{Discussion}

The bulk and boundary scattering in the context of the $\mathfrak{sl}(3)$
twisted Yangian is studied. The analysis in based on the Bethe ansatz
methodology. In particular, the thermodynamic limit of the associated
Bethe ansatz equations is studied and the ground state and excitations
are determined. The scattering among the particle-like excitations gives
rise to a factorized form expressed explicitly as a product of the
soliton$-$soliton times the soliton$-$anti-soliton scattering amplitude of
the bulk $\mathfrak{sl}(3)$ case. Moreover, the interaction of the excitation with
the boundary is studied and the corresponding boundary scattering
amplitude is derived. Note that we have considered here the simplest
boundary matrices i.e. $K^{\pm} \propto {\mathbb I}$ ($K^{\pm}(\l) = k^{\pm}(\l)\ {\mathbb I}$). One of the key
points in this investigation together with the study of the boundary
scattering is the formulation of the suitable quantization condition
compatible with the underlying algebraic setting as well as the
corresponding physical interpretation. This is also confirmed by the fact
that the bulk scattering factorizes into the product of the soliton-soliton
and soliton--antisoliton scattering amplitudes.

It is worth pointing out that in the particular case under study as well as
for the generic $\mathfrak{sl}(2n+1)$ case the Bethe ansatz equations
are similar to the $\mathfrak{osp}(1|2n)$ case, whereas in the
$\mathfrak{sl}(2n)$
case they are a bit modified. In any case, the next natural step is to
generalize these computations for the $\mathfrak{sl}(n)$ case.
Furthermore, the study of defects within the context of the twisted
Yangian is a very interesting direction to pursue. Hopefully, the
aforementioned issues will be addressed in a forthcoming publication.


\subsection*{Acknowledgments}
A.D. wishes to thank University of Cergy-Pontoise, where part of this work was completed, for kind hospitality.
We thank the referees for their helpful comments and suggestions on the presentation of our results.


\begin{thebibliography}{1}

\bibitem{cherednik} I.V. Cherednik, Theor. Math. Phys. {\bf 61} (1984) 977.

\bibitem{sklyanin} E.K. Sklyanin, J. Phys. {\bf A21} (1988) 2375.

\bibitem{fre-mai}
L. Freidel, J. M. Maillet, Phys. Lett {\bf B262} (1991) 268.

\bibitem{kulish}
P.P. Kulish and E.K Sklyanin,  J. Phys. {\bf A25} (1992) 5963, [hep-th/9209054 ].

\bibitem{more}
A.I. Molev and E. Ragoucy, Rev. Math. Phys. 14 (2002) 317, [arXiv:math/0107213];\\
E. Ragoucy and S. Belliard, J. Phys. {\bf A42} (2009) 205203, [arXiv:0902.0321, math-ph].

\bibitem{Doikou:2000yw}
  A.~Doikou,
J.\ Phys.\ A {\bf 33} (2000) 8797,
  [hep-th/0006197].

  \bibitem{Arnaudon:2004sd}
  D.~Arnaudon, J.~Avan, N.~Crampe, A.~Doikou, L.~Frappat and E.~Ragoucy,
  J.\ Stat.\ Mech.\  {\bf 0408} (2004) P08005,
  [math-ph/0406021].


\bibitem{twisted} G.I. Olshanski, Twisted Yangians and infinite-dimensional classical Lie algebras in ‘Quantum
Groups’ (P.P. Kulish, Ed.), Lecture notes in Math. 1510, Springer (1992) pp 103;\\
A. Molev, M. Nazarov and G.I. Olshanski, Russ. Math. Surveys 51:2 (1996) 205.

 \bibitem{corrigan}
  E. Corrigan, P.E. Dorey, R.H. Rietdijk and R. Sasaki, Phys. Lett. {\bf B333} (1994) 83, [hep-th/9404108].


\bibitem{doikou-affine} A. Doikou, JHEP 0805:091 (2008), [arXiv:0803.0943, hep-th ];\\
J. Avan and A. Doikou,  Nucl. Phys. {\bf B821} (2009) 481, [arXiv:0809.2734, hep-th].

\bibitem{fad}
L. Faddeev, E. Sklyanin and L. Takhtajan, Theor. Math. Phys. {\bf 40} (1980) 688;\\
N. Yu. Reshethikhin, L. Takhtajan and L.D. Faddeev, Len. Math. J. {\bf 1} (1990) 193.


\bibitem{Arnaudon:2003zw}
  D.~Arnaudon, J.~Avan, N.~Crampe, A.~Doikou, L.~Frappat and E.~Ragoucy,
  Nucl.\ Phys.\ B {\bf 687} (2004) 257,
  [math-ph/0310042].



\bibitem{ADNK}
 J. Avan, A. Doikou and N. Karaiskos, {\it The $\mathfrak{sl}({\cal N})$ twisted Yangian: bulk-boundary scattering $\&$ defects}, [arXiv:1412.6480, hep-th].

\bibitem{Yang:1967bm}
  C.~N.~Yang,
  Phys.\ Rev.\ Lett.\  {\bf 19} (1967) 1312.

\bibitem{Korepin:1979hg}
  V.~E.~Korepin,
  Commun.\ Math.\ Phys.\  {\bf 76} (1980) 165.

\bibitem{Andrei:1983cb}
  N.~Andrei and C.~Destri,
  Nucl.\ Phys.\ B {\bf 231} (1984) 445.


  \bibitem{Grisaru:1994ru}
  M.~T.~Grisaru, L.~Mezincescu and R.~I.~Nepomechie,
  J.\ Phys.\ A {\bf 28} (1995) 1027,
  [hep-th/9407089];\\
  A.~Doikou, L.~Mezincescu and R.~I.~Nepomechie,
  J.\ Phys.\ A {\bf 30} (1997) L507,
  [hep-th/9705187].


\bibitem{Suth1975}
 B. Sutherland, Phys. Rev. B\textbf{12} (1975) 3795.

\bibitem{doikou-nepo}
A. Doikou and R.I. Nepomechie, Nucl. Phys. {\bf B521} (1998) 547, [hep-th/9803118].


\end{thebibliography}
\end{document}